\documentclass[12pt]{article}
\usepackage{amsmath,amsthm}
\usepackage{eufrak}
\usepackage{pb-diagram,lamsarrow,pb-lams}

\newcommand{\ar}{\renewcommand{\arraystretch}{1}} 
\DeclareMathAlphabet{\bb}{U}{msb}{m}{n} \gdef\C{\bb C}     \gdef\R{\bb R}

 \DeclareMathOperator{\spin}{{\bf
Spin}} 
\DeclareMathOperator{\fD}{\mathfrak{D}}

\DeclareMathOperator{\Sym}{Sym}

 \DeclareMathOperator{\sign}{sign}
\DeclareMathOperator{\SO}{SO}
\DeclareMathOperator{\SL}{SL} \DeclareMathOperator{\SU}{SU}
\DeclareMathOperator{\QU}{QU}

\newcommand{\scr}{\scriptstyle}

\newcommand{\re}{\mbox{\rm Re}\,}
\newcommand{\im}{\mbox{\rm Im}\,}

\newcommand{\cM}{{\cal M}}

\newcommand{\sA}{{\sf A}}
\newcommand{\sB}{{\sf B}}

\newcommand{\sX}{{\sf X}}
\newcommand{\sY}{{\sf Y}}

\newcommand{\bx}{{\bf x}}
\newcommand{\by}{{\bf y}}

\newcommand{\fM}{\mathfrak{M}}

\newcommand{\fP}{\mathfrak{P}}
\newcommand{\fL}{\mathfrak{L}}

\newcommand{\fS}{\mathfrak{S}}

\newcommand{\fg}{\mathfrak{g}}

\newcommand{\cl}{C\kern -0.2em \ell}

\newcommand{\hypergeom}[5]{\mbox{$
_#1 F_#2\left. \!\! \left( \!\!\ar
\begin{array}{c}
\multicolumn{1}{c}{\begin{array}{c} #3
\end{array}}\\[1mm]
\multicolumn{1}{c}{\begin{array}{c} #4
\end{array}}\end{array}
\!\! \right|\displaystyle{#5}\right) $} }
\newcommand{\ld}{\left[}
\newcommand{\rd}{\right]}

\newcommand{\tg}{\tan}
\newcommand{\ch}{\cosh}
\newcommand{\sh}{\sinh}
\newcommand{\tnh}{\tanh}

\newcommand{\ctg}{\cot}
\newtheorem{theorem}{Theorem}
\begin{document}
\title{Relativistic Spherical Functions on the Lorentz Group}
\author{V. V. Varlamov\\
{\small\it Department of Mathematics, Siberia State University of Industry}\\
{\small\it Kirova 42, Novokuznetsk 654007, Russia}}
\date{}
\maketitle
\begin{abstract}
Matrix elements of irreducible representations of the Lorentz group
are calculated on the basis of complex angular momentum. It is shown
that Laplace-Beltrami operators, defined in this basis, give rise to
Fuchsian differential equations. An explicit form of the matrix
elements of the Lorentz group has been found via the addition
theorem for generalized spherical functions. Different expressions
of the matrix elements are given in terms of hypergeometric
functions both for finite-dimensional and unitary representations of
the principal and supplementary series of the Lorentz group.
\end{abstract}
\section{Introduction}
As is known, an expansion problem of relativistic amplitudes
requires the most simple form for the matrix elements of irreducible
representations of the Lorentz group. Matrix elements of this group
are studied for a long time by many authors. So, in 1956, Dolginov
\cite{Dol56} (see also \cite{DT59,DM59,Esk59}) considered an
analytic continuation of the Fock four-dimensional spherical
functions (four-dimensional spherical functions of an euclidean
space was introduced by Fock \cite{Foc35} for the solution of the
hydrogen atom problem in momentum representation). Basis functions,
called in the works \cite{Dol56,DT59,DM59} as {\it relativistic
spherical functions}, depend on angles of the radius-vector in the
four-dimensional spacetime. It should be noted that
Dolginov-Toptygin relativistic spherical functions present the most
degenerate form of the matrix elements of the Lorentz group.
Different realizations of these elements were studied in the works
\cite{Esk61,Gol61,Str65,Str67,Str68,ST67,VD67,Kol70}. The most
complete form of the matrix elements of the Lorentz group was given
in the works \cite{Str65,Str67} within the Gel'fand-Naimark basis
\cite{GMS,Nai58}. However, matrix elements in the Str\"{o}m form,
and also in the Sciarrino-Toller form \cite{ST67}, are very
complicate and cumbersome. Smorodinsky and Huszar
\cite{SH70,Hus85,Hus88} found more simple and direct method for
definition of the matrix elements of the Lorentz group by means of a
complexification of the three-dimensional rotation group and
solution of the equation on eigenvalues of the Casimir operators
(see also \cite{AG64}).

In the present work matrix elements of irreducible representations
of the Lorentz group are found on the basis of complex angular
momentum ($\SU(2)\otimes\SU(2)$-basis). It is shown that
Laplace-Beltrami operators, defined in this basis, lead to Fuchsian
differential equations which can be reduced to hypergeometric
equations. An explicit form of the matrix elements has been found
via the addition theorem for generalized spherical functions, where
the functions $P^l_{mn}$ and $\fP^l_{mn}$ are components. As is
known \cite{GS52}, the matrix elements of $\SU(2)$ are defined by
the functions $P^l_{mn}$, and matrix elements of the group $\QU(2)$
of quasiunitary matrices of the second order, which is isomorphic to
the group $\SL(2,\R)$ \footnote{Other designation of this group is
$\SU(1,1)$ known also as three-dimensional Lorentz group,
representations of which was studied by Bargmann \cite{Bar47}.}, are
expressed via the functions $\fP^l_{mn}$. The groups $\SU(2)$ and
$SU(1,1)$ are real forms of the group $\SL(2,\C)$. The factorization
of the matrix elements of $\SL(2,\C)$ with respect to the subgroups
$\SU(2)$ and $\SU(1,1)$ allows us to express these elements via the
product of two hypergeometric functions both for finite-dimensional
and unitary representations of the principal and supplementary
series of the Lorentz group (it should be noted that matrix elements
in Str\"{o}m form are expressed via the product of three
hypergeometric functions). On the other hand, matrix elements of the
Lorentz group play an essential role in quantum field theory on the
Poincar\'{e} group \cite{Kih70,Tol96,Dre97,GS01,Var03c,Var03d},
where the field operators are expressed via generalized Fourier
integrals (it leads to harmonic analysis on the homogeneous spaces).
Solutions of relativistic wave equations are reduced also to
expansions in relativistic spherical functions \cite{AAV69,Var03}.
Moreover, the Biedenharn type relativistic wavefunctions
\cite{BBTD88} are defined completely in this framework
\cite{Var03c,Var03d}.

\section{Relativistic Spherical Functions}
As is known, the group $\spin_+(1,3)\simeq\SL(2,\C)$ is an universal
covering of the proper orthochronous Lorentz group $\SO_0(1,3)$. The
group $\SL(2,\C)$ of all complex matrices
\[\ar
\fg=
\begin{pmatrix}
\alpha & \beta\\
\gamma & \delta
\end{pmatrix}
\]
of 2-nd order with the determinant $\alpha\delta-\gamma\beta=1$, is
a {\it complexification} of the group $\SU(2)$. The group $\SU(2)$
is one of the real forms of $\SL(2,\C)$. The transition from
$\SU(2)$ to $\SL(2,\C)$ is realized via the complexification of
three real parameters $\varphi,\,\theta,\,\psi$ (Euler angles) of
$\SU(2)$. Let $\theta^c=\theta-i\tau$,
$\varphi^c=\varphi-i\epsilon$, $\psi^c=\psi-i\varepsilon$ be complex
Euler angles, where
\begin{equation}\label{CEA}
{\renewcommand{\arraystretch}{1.05}
\begin{array}{ccccc}
0 &\leq&\re\theta^c=\theta& \leq& \pi,\\
0 &\leq&\re\varphi^c=\varphi& <&2\pi,\\
-2\pi&\leq&\re\psi^c=\psi&<&2\pi,
\end{array}\quad\quad
\begin{array}{ccccc}
-\infty &<&\im\theta^c=\tau&<&+\infty,\\
-\infty&<&\im\varphi^c=\epsilon&<&+\infty,\\
-\infty&<&\im\psi^c=\varepsilon&<&+\infty.
\end{array}}
\end{equation}
Infinitesimal operators $\sA_i$ and $\sB_i$ of the group $\SL(2,\C)$
form a basis of the Lie algebra $\mathfrak{sl}(2,\C)$ and satisfy
the relations
\begin{equation}\label{Com1}
\left.\begin{array}{lll} \ld\sA_1,\sA_2\rd=\sA_3, &
\ld\sA_2,\sA_3\rd=\sA_1, &
\ld\sA_3,\sA_1\rd=\sA_2,\\[0.1cm]
\ld\sB_1,\sB_2\rd=-\sA_3, & \ld\sB_2,\sB_3\rd=-\sA_1, &
\ld\sB_3,\sB_1\rd=-\sA_2,\\[0.1cm]
\ld\sA_1,\sB_1\rd=0, & \ld\sA_2,\sB_2\rd=0, &
\ld\sA_3,\sB_3\rd=0,\\[0.1cm]
\ld\sA_1,\sB_2\rd=\sB_3, & \ld\sA_1,\sB_3\rd=-\sB_2, & \\[0.1cm]
\ld\sA_2,\sB_3\rd=\sB_1, & \ld\sA_2,\sB_1\rd=-\sB_3, & \\[0.1cm]
\ld\sA_3,\sB_1\rd=\sB_2, & \ld\sA_3,\sB_2\rd=-\sB_1. &
\end{array}\right\}
\end{equation}
Let us consider the operators
\begin{gather}
\sX_l=\frac{1}{2}i(\sA_l+i\sB_l),\quad\sY_l=\frac{1}{2}i(\sA_l-i\sB_l),
\label{SL25}\\
(l=1,2,3).\nonumber
\end{gather}
Using the relations (\ref{Com1}), we obtain
\begin{equation}\label{Com2}
\ld\sX_k,\sX_l\rd=i\varepsilon_{klm}\sX_m,\quad
\ld\sY_l,\sY_m\rd=i\varepsilon_{lmn}\sY_n,\quad \ld\sX_l,\sY_m\rd=0.
\end{equation}
Further, introducing generators of the form
\begin{equation}\label{SL26}
\left.\begin{array}{cc}
\sX_+=\sX_1+i\sX_2, & \sX_-=\sX_1-i\sX_2,\\[0.1cm]
\sY_+=\sY_1+i\sY_2, & \sY_-=\sY_1-i\sY_2,
\end{array}\right\}
\end{equation}
we see that in virtue of commutativity of the relations (\ref{Com2})
a space of an irreducible finite--dimensional representation of the
group $\SL(2,\C)$ can be spanned on the totality of
$(2l+1)(2\dot{l}+1)$ basis vectors $\mid
l,m;\dot{l},\dot{m}\rangle$, where $l,m,\dot{l},\dot{m}$ are integer
or half--integer numbers, $-l\leq m\leq l$, $-\dot{l}\leq
\dot{m}\leq \dot{l}$. Therefore,
\begin{eqnarray}
&&\sX_-\mid l,m;\dot{l},\dot{m}\rangle= \sqrt{(l+m)(l-m+1)}\mid
l,m-1,\dot{l},\dot{m}\rangle
\;\;(m>-l),\nonumber\\
&&\sX_+\mid l,m;\dot{l},\dot{m}\rangle= \sqrt{(l-m)(l+m+1)}\mid
l,m+1;\dot{l},\dot{m}\rangle
\;\;(m<l),\nonumber\\
&&\sX_3\mid l,m;\dot{l},\dot{m}\rangle=
m\mid l,m;\dot{l},\dot{m}\rangle,\nonumber\\
&&\sY_-\mid l,m;\dot{l},\dot{m}\rangle=
\sqrt{(\dot{l}+\dot{m})(\dot{l}-\dot{m}+1)}\mid
l,m;\dot{l},\dot{m}-1
\rangle\;\;(\dot{m}>-\dot{l}),\nonumber\\
&&\sY_+\mid l,m;\dot{l},\dot{m}\rangle=
\sqrt{(\dot{l}-\dot{m})(\dot{l}+\dot{m}+1)}\mid
l,m;\dot{l},\dot{m}+1
\rangle\;\;(\dot{m}<\dot{l}),\nonumber\\
&&\sY_3\mid l,m;\dot{l},\dot{m}\rangle= \dot{m}\mid
l,m;\dot{l},\dot{m}\rangle.\label{Waerden}
\end{eqnarray}
From relations (\ref{Com2}), it follows that each of the sets of
infinitesimal operators $\sX$ and $\sY$ generates the group $\SU(2)$
and these two groups commute with each other. Thus, from the
relations (\ref{Com2}) and (\ref{Waerden}) it follows that the group
$\SL(2,\C)$, in essence, is equivalent locally to the group
$\SU(2)\otimes\SU(2)$. The basis (\ref{Waerden}) was first
introduced by Van der Waerden in \cite{Wa32}

On the group $\SL(2,\C)$ there exist the following Laplace-Beltrami
operators:
\begin{eqnarray}
\sX^2&=&\sX^2_1+\sX^2_2+\sX^2_3=\frac{1}{4}(\sA^2-\sB^2+2i\sA\sB),\nonumber\\
\sY^2&=&\sY^2_1+\sY^2_2+\sY^2_3=
\frac{1}{4}(\widetilde{\sA}^2-\widetilde{\sB}^2-
2i\widetilde{\sA}\widetilde{\sB}).\label{KO}
\end{eqnarray}
At this point, we see that operators (\ref{KO}) contain the well
known Casimir operators $\sA^2-\sB^2$, $\sA\sB$ of the Lorentz
group. Using expressions (\ref{CEA}), we obtain a Euler
parametrization of the Laplace-Beltrami operators,
\begin{eqnarray}
\sX^2&=&\frac{\partial^2}{\partial\theta^c{}^2}+
\ctg\theta^c\frac{\partial}{\partial\theta^c}+\frac{1}{\sin^2\theta^c}\left[
\frac{\partial^2}{\partial\varphi^c{}^2}-
2\cos\theta^c\frac{\partial}{\partial\varphi^c}
\frac{\partial}{\partial\psi^c}+
\frac{\partial^2}{\partial\psi^c{}^2}\right],\nonumber\\
\sY^2&=&\frac{\partial^2}{\partial\dot{\theta}^c{}^2}+
\ctg\dot{\theta}^c\frac{\partial}{\partial\dot{\theta}^c}+
\frac{1}{\sin^2\dot{\theta}^c}\left[
\frac{\partial^2}{\partial\dot{\varphi}^c{}^2}-
2\cos\dot{\theta}^c\frac{\partial}{\partial\dot{\varphi}^c}
\frac{\partial}{\partial\dot{\psi}^c}+
\frac{\partial^2}{\partial\dot{\psi}^c{}^2}\right].\label{KO2}
\end{eqnarray}
Here $\dot{\theta}^c=\theta+i\tau$,
$\dot{\varphi}^c=\varphi+i\epsilon$,
$\dot{\psi}^c=\psi+i\varepsilon$ are complex conjugate Euler angles.

Matrix elements $t^{l}_{mn}(\fg)=
\fM^{l}_{mn}(\varphi^c,\theta^c,\psi^c)$ of irreducible
representations of the group $\SL(2,\C)$ are eigenfunctions of the
operators (\ref{KO2}),
\begin{eqnarray}
\left[\sX^2+l(l+1)\right]\fM^{l}_{mn}(\varphi^c,\theta^c,\psi^c)&=&0,\nonumber\\
\left[\sY^2+\dot{l}(\dot{l}+1)\right]\fM^{\dot{l}}_{\dot{m}\dot{n}}
(\dot{\varphi}^c,\dot{\theta}^c,\dot{\psi}^c)&=&0,\label{EQ}
\end{eqnarray}
where
\begin{eqnarray}
\fM^{l}_{mn}(\fg)&=& e^{-i(m\varphi^c+n\psi^c)}Z^{l}_{mn}
(\cos\theta^c),\nonumber\\
\fM^{\dot{l}}_{\dot{m}\dot{n}}(\fg)&=&e^{i(\dot{m}\dot{\varphi}^c+
\dot{n}\dot{\psi}^c)}Z^{\dot{l}}_{\dot{m}\dot{n}}(\cos\dot{\theta}^c).
\label{HF3'}
\end{eqnarray}
Here $\fM^l_{mn}(\fg)$ are general matrix elements of the
representations of $\SO_0(1,3)$, and $Z^l_{mn}(\cos\theta^c)$ are
{\it hyperspherical functions}. Substituting the functions
(\ref{HF3'}) into (\ref{EQ}) and taking into account the operators
(\ref{KO2}) and substitutions $z=\cos\theta^c$,
$\overset{\ast}{z}=\cos\dot{\theta}^c$, we arrive at the following
differential equations:
\begin{eqnarray}
\left[(1-z^2)\frac{d^2}{dz^2}-2z\frac{d}{dz}-
\frac{m^2+n^2-2mnz}{1-z^2}+l(l+1)\right]
Z^{l}_{mn}(z)&=&0,\label{Leg1}\\
\left[(1-\overset{\ast}{z}{}^2)\frac{d^2}{d\overset{\ast}{z}{}^2}-
2\overset{\ast}{z}\frac{d}{d\overset{\ast}{z}}-
\frac{\dot{m}^2+\dot{n}^2-2\dot{m}\dot{n}\overset{\ast}{z}}
{1-\overset{\ast}{z}{}^2}+\dot{l}(\dot{l}+1)\right]
Z^{\dot{l}}_{\dot{m}\dot{n}}(\overset{\ast}{z})&=&0.\label{Leg2}
\end{eqnarray}
The latter equations have three singular points $-1$, $+1$,
$\infty$. The equations (\ref{Leg1}), (\ref{Leg2}) are Fuchsian
equations. Indeed, denoting $w(z)=Z^l_{mn}(z)$, we write the
equation (\ref{Leg1}) in the form
\begin{equation}\label{Fux}
\frac{d^2w(z)}{dz^2}-p(z)\frac{dw(z)}{dz}+q(z)w(z)=0,
\end{equation}
where
\[
p(z)=\frac{2z}{(1-z)(1+z)},\quad
q(z)=\frac{l(l+1)(1-z^2)-m^2-n^2+2mnz}{(1-z)^2(1+z)^2}.
\]
Let us find solutions of (\ref{Leg1}). Applying the substitution
\[
t=\frac{1-z}{2},\quad
w(z)=t^{\frac{|m-n|}{2}}(1-t)^{\frac{|m+n|}{2}}v(t),\nonumber
\]
we arrive at hypergeometric equation
\begin{equation}\label{Hyper}
t(1-t)\frac{d^2v}{dt^2}+[c-(a+b+1)t]\frac{dv}{dt}-abv(t)=0,
\end{equation}
where
\begin{eqnarray}
a&=&l+1+\frac{1}{2}(|m-n|+|m+n|),\nonumber\\
b&=&-l+\frac{1}{2}(|m-n|+|m+n|),\nonumber\\
c&=&|m-n|+1.\nonumber
\end{eqnarray}
Therefore, a solution of (\ref{Hyper}) is
\[
v(t)=C_1\hypergeom{2}{1}{a,b}{c}{t}+C_2t^{1-c}
\hypergeom{2}{1}{b-c+1,a-c+1}{2-c}{t}.
\]
Coming back to initial variable, we obtain
\begin{multline}
w(z)=C_1\left(\frac{1-z}{2}\right)^{\frac{|m-n|}{2}}
\left(\frac{1+z}{2}\right)^{\frac{|m+n|}{2}}\times\\
\times\hypergeom{2}{1}{l+1+\frac{1}{2}(|m-n|+|m+n|),-l+\frac{1}{2}(|m-n|+|m+n|)}
{|m-n|+1}{\frac{1-z}{2}}+\\
+C_2\left(\frac{1-z}{2}\right)^{-\frac{|m-n|}{2}}
\left(\frac{1+z}{2}\right)^{\frac{|m+n|}{2}}\times\\
\times\hypergeom{2}{1}{-l+\frac{1}{2}(|m+n|-|m-n|),l+1+\frac{1}{2}(|m+n|-|m-n|)}
{1-|m-n|}{\frac{1-z}{2}}. \label{Sol1}
\end{multline}
Carrying out the analogous calculations for the equation
(\ref{Leg2}), we find that
\begin{multline}
w(\overset{\ast}{z})=
C_1\left(\frac{1-\overset{\ast}{z}}{2}\right)^{\frac{|\dot{m}-\dot{n}|}{2}}
\left(\frac{1+\overset{\ast}{z}}{2}\right)^{\frac{|\dot{m}+\dot{n}|}{2}}\times\\
\times\hypergeom{2}{1}{l+1+\frac{1}{2}(|\dot{m}-\dot{n}|+|\dot{m}+\dot{n}|),-l+
\frac{1}{2}(|\dot{m}-\dot{n}|+|\dot{m}+\dot{n}|)}
{|\dot{m}-\dot{n}|+1}{\frac{1-\overset{\ast}{z}}{2}}+\\
+C_2\left(\frac{1-\overset{\ast}{z}}{2}\right)^{-\frac{|\dot{m}-\dot{n}|}{2}}
\left(\frac{1+\overset{\ast}{z}}{2}\right)^{\frac{|\dot{m}+\dot{n}|}{2}}\times\\
\times\hypergeom{2}{1}{-l+\frac{1}{2}(|\dot{m}+\dot{n}|-|\dot{m}-\dot{n}|),
l+1+\frac{1}{2}(|\dot{m}+\dot{n}|-|\dot{m}-\dot{n}|)}
{1-|\dot{m}-\dot{n}|}{\frac{1-\overset{\ast}{z}}{2}}. \label{Sol2}
\end{multline}
\begin{sloppypar}
As follows from (\ref{Sol1}) and (\ref{Sol2}), the functions
$Z^l_{mn}$ and $Z^{\dot{l}}_{\dot{m}\dot{n}}$ are expressed via the
hypergeometric function. In virtue of the full development of the
theory of hypergeometric functions, the representations (\ref{Sol1})
and (\ref{Sol2}) are the most useful. Indeed, from (\ref{Sol1}) it
follows that the function $Z^l_{mn}$ can be represented by the
following particular solution:
\end{sloppypar}
\begin{multline}
Z^l_{mn}(\cos\theta^c)=C_1\sin^{|m-n|}\frac{\theta^c}{2}
\cos^{|m+n|}\frac{\theta^c}{2}\times\\
\times\hypergeom{2}{1}{l+1+\frac{1}{2}(|m-n|+|m+n|),-l+\frac{1}{2}(|m-n|+|m+n|)}
{|m-n|+1}{\sin^2\frac{\theta^c}{2}}.\label{Hyper2}
\end{multline}

Let us give now a general definition for spherical functions on the
group $G$. Let $T(g)$ be an irreducible representation of the group
$G$ in the space $L$ and let $H$ be a subgroup of $G$. The vector
$\boldsymbol{\xi}$ in the space $L$ is called {\it an invariant with
respect to the subgroup} $H$ if for all $h\in H$ the equality
$T(h)\boldsymbol{\xi}=\boldsymbol{\xi}$ holds. The representation
$T(g)$ is called {\it a representation of the class one with respect
to the subgroup} $H$ if in its space there are non-null vectors
which are invariant with respect to $H$. At this point, a
contraction of $T(g)$ onto its subgroup $H$ is unitary:
\[
(T(h)\boldsymbol{\xi}_1,T(h)\boldsymbol{\xi}_2)=(\boldsymbol{\xi}_1,
\boldsymbol{\xi}_2).
\]
Hence it follows that a function
\[
f(g)=(T(g)\boldsymbol{\eta},\boldsymbol{\xi})
\]
corresponds the each vector $\boldsymbol{\eta}\in L$. $f(g)$ are
called {\it spherical functions of the representation $T(g)$ with
respect to $H$}.

Spherical functions can be considered as functions on homogeneous
spaces $\cM=G/H$. In its turn,
a homogeneous space $\cM$ of the group $G$ has the following properties:\\
a) It is a topological space on which the group $G$ acts
continuously, that is, let $y$ be a point in $\cM$, then $gy$ is
defined and is again
a point in $\cM$ ($g\in G$).\\
b) This action is transitive, that is, for any two points $y_1$ and
$y_2$ in $\cM$ it is always possible to find a group element $g\in
G$
such that $y_2=gy_1$.\\
There is a one-to-one correspondence between the homogeneous spaces
of $G$ and the coset spaces of $G$. Let $H_0$ be a maximal subgroup
of $G$ which leaves the point $y_0$ invariant, $hy_0=y_0$, $h\in
H_0$, then $H_0$ is called the stabilizer of $y_0$. Representing now
any group element of $G$ in the form $g=g_ch$, where $h\in H_0$ and
$g_c\in G/H_0$, we see that, by virtue of the transitivity property,
any point $y\in\cM$ can be given by $y=g_chy_0=g_cy$. Hence it
follows that the elements $g_c$ of the coset space give a
parametrization of $\cM$. The mapping $\cM\leftrightarrow G/H_0$ is
continuous since the group multiplication is continuous and the
action on $\cM$ is continuous by definition. The stabilizers $H$ and
$H_0$ of two different points $y$ and $y_0$ are conjugate, since
from $H_0g_0=g_0$, $y_0=g^{-1}y$, it follows that $gH_0g^{-1}y=y$,
that is, $H=gH_0g^{-1}$.

Coming back to the Lorentz group $G=\SO_0(1,3)$, we see that there
are the following homogeneous spaces of $\SO_0(1,3)$ depending on
the stabilizer $H$. First of all, when $H=0$ the homogeneous space
$\cM_6$ coincides with {\it a group manifold} $\fL_6$ of
$\SO_0(1,3)$. Therefore, $\fL_6$ is a maximal homogeneous space of
the Lorentz group. Further, when $H=\Omega^c_\psi$, where
$\Omega^c_\psi$ is a group of diagonal matrices $\begin{pmatrix}
e^{\frac{i\psi^c}{2}} & 0\\
0 & e^{-\frac{i\psi^c}{2}}
\end{pmatrix}$, the homogeneous space $\cM_4$ coincides with a
two-dimensional complex sphere $S^c_2$,
$\cM_4=S^c_2\sim\SL(2,\C)/\Omega^c_\psi$. The sphere $S^c_2$ can be
constructed from the quantities $z_k=x_k+iy_k$,
$\overset{\ast}{z}_k=x_k-iy_k$ $(k=1,2,3)$ as follows:
\begin{equation}\label{CS}
S^c_2:\;z^2_1+z^2_2+z^2_3=\bx^2-\by^2+2i\bx\by=r^2.
\end{equation}
The complex conjugate (dual) sphere $\dot{S}^c_2$ is
\begin{equation}\label{DS}
\dot{S}^c_2:\;\overset{\ast}{z}_1{}^2+\overset{\ast}{z}_2{}^2+
\overset{\ast}{z}_3{}^2=\bx^2-\by^2-2i\bx\by=\overset{\ast}{r}{}^2.
\end{equation}
The following homogeneous space $\cM_3$ we obtain when the
stabilizer $H$ coincides with a maximal compact subgroup $K=\SO(3)$
of $\SO_0(1,3)$. In this case we have a three-dimensional
two-sheeted hyperboloid
$\cM_3=H_3\sim\SO_0(1,3)/\SO(3)\simeq\SL(2,\C)/\SU(2)$, defined by
the equation
\[
H_3=\{x\in\R^{1,3}|\ld x,x\rd=1\}.
\]
In the case $\ld x,x\rd=0$ we arrive at a cone $C_3$ which can be
considered also as a homogeneous space of $\SO_0(1,3)$. Usually,
only the upper sheets $H^+_3$ and $C^+_3$ are considered in
applications.

Finally, a minimal homogeneous space $\cM_2$ of $\SO_0(1,3)$ is a
two-dimensional real sphere $S_2\sim\SO(3)/\SO(2)$. In contrast to
the previous homogeneous spaces, the sphere $S_2$ coincides with a
quotient space $\SO_0(1,3)/P$, where $P$ is a minimal parabolic
subgroup of $\SO_0(1,3)$. From the Iwasawa decompositions
$\SO_0(1,3)=KNA$ and $P=MNA$, where $M=\SO(2)$, $N$ and $A$ are
nilpotent and commutative subgroups of $\SO_0(1,3)$, it follows that
$\SO_0(1,3)/P=KNA/MNA\sim K/M\sim\SO(3)/\SO(2)$.

Taking into account the list of homogeneous spaces of $\SO_0(1,3)$,
we introduce now the following types of spherical functions $f(\fg)$
on the Lorentz group:
\begin{itemize}
\item $f(\fg)=\fM^l_{mn}(\fg)$. This function is defined on the
group manifold $\fL_6$ of $\SO_0(1,3)$. It is the most general
spherical function on the group $\SO_0(1,3)$. In this case $f(\fg)$
depends on all the six parameters of $\SO_0(1,3)$ and for that
reason it should be called as {\it a function on the Lorentz group}.
An explicit form of $\fM^l_{mn}(\fg)$ (respectively
$\fM^{\dot{l}}_{\dot{m}\dot{n}}(\fg)$) for finite-dimensional
representations and of $\fM^{-\frac{1}{2}+i\rho}_{mn}(\fg)$ (resp.
$\fM^{-\frac{1}{2}-i\rho}_{\dot{m}\dot{n}}(\fg)$) for
infinite-dimensional representations of $\SO_0(1,3)$ will be given
in the sections 3 and 4, respectively.
\item $f(\varphi^c,\theta^c)=\fM^m_l(\varphi^c,\theta^c,0)$.
This function is defined on the homogeneous space
$\cM_4=S^c_2\sim\SO_0(1,3)/\Omega^c_\psi$, that is, on the surface
of the two-dimensional complex sphere $S^c_2$. The function
$\fM^m_l(\varphi^c,\theta^c,0)$ is a relativistic analogue of the
usual spherical function $Y^m_l(\varphi,\theta)$ defined on the
surface of the real two-sphere $S_2$. In its turn, the function
$f(\dot{\varphi}^c,\dot{\theta}^c)=
\fM^{\dot{m}}_{\dot{l}}(\dot{\varphi}^c,\dot{\theta}^c,0)$ is
defined on the surface of the dual sphere $\dot{S}^c_2$. General
solutions of relativistic wave equations have been found via an
expansion in spherical functions $f(\varphi^c,\theta^c)$
\cite{Var03}. An explicit form of the functions
$\fM^m_l(\varphi^c,\theta^c,0)$
($\fM^{\dot{m}}_{\dot{l}}(\dot{\varphi}^c,\dot{\theta}^c,0)$) and
$\fM^m_{-\frac{1}{2}+i\rho}(\varphi^c,\theta^c,0)$
($\fM^{\dot{m}}_{-\frac{1}{2}-i\rho}(\dot{\varphi}^c,\dot{\theta}^c,0)$)
will be given in the section 3 and 4.
\item
$f(\epsilon,\tau,\varepsilon)=e^{-im\epsilon}\fP^l_{mn}(\cosh\tau)
e^{-in\varepsilon}$. This function is defined on the homogeneous
space $\cM_3=H^+_3\sim\SO_0(1,3)/\SO(3)$, that is, on the upper
sheet of the hyperboloid $x^2_0-x^2_1-x^2_2-x^2_3=1$. In essence, we
come here to representations of $\SO_0(1,3)$ restricted to the
subgroup $\SU(1,1)$ \cite{ST67,Kol70}.
\item $f(\varphi,\theta,\psi)=e^{-im\varphi}P^l_{mn}(\cos\theta)e^{-in\psi}$.
This function is defined on the homogeneous space
$\cM_2=S_2\sim\SO(3)/\SO(2)$, that is, on the surface of the
two-dimensional real sphere $S_2$. We come here to the most
degenerate representations of $\SO_0(1,3)$ restricted to the
subgroup $\SU(2)$.
\end{itemize}
We see that only first two functions $f(\fg)$ and
$f(\varphi^c,\theta^c)$ can be considered as functions on the
Lorentz group $\SO_0(1,3)$; other two functions
$f(\epsilon,\tau,\varepsilon)$ and $f(\varphi,\theta,\psi)$ present
degenerate cases corresponding to the subgroups $\SU(1,1)$ and
$\SU(2)$. For that reason the functions $f(\fg)$ and
$f(\varphi^c,\theta^c)$ should be called {\it relativistic spherical
functions on the Lorentz group}.

\section{Hyperspherical Functions and Addition Theorem for
Generalized Spherical Functions} In this section we will find
expressions for the matrix elements (relativistic spherical
functions) containing explicitly all six parameters of the Lorentz
group. Moreover, such a form of the matrix elements to be the most
suitable for forthcoming tasks of harmonic analysis on the Lorentz
and Poincar\'{e} groups.

As is known,  the groups $\SU(2)$ and $\SU(1,1)\simeq \SL(2,\R)$ are
real forms of $\SL(2,\C)$. As a direct consequence of this, a
structure of the matrix elements of these groups is very similar
with a corresponding structure of matrix elements for the group
$\SL(2,\C)$. Indeed, matrix elements of irreducible representations
of $\SU(2)$ have the form \cite{Vil65,VK90}
\[
t^l_{mn}(u)=e^{-im\varphi}P^l_{mn}(\cos\theta)e^{-in\psi},
\]
where
\begin{multline}
P^l_{mn}(\cos\theta)=i^{m-n}
\sqrt{\frac{\Gamma(l-m+1)\Gamma(l-n+1)}{\Gamma(l+m+1)\Gamma(l+n+1)}}\times\\
\times\cos^{m+n}\frac{\theta}{2}\sin^{m-n}\frac{\theta}{2}\sum^{l-m}_{t=0}
\frac{(-1)^t\Gamma(l+m+t+1)}{\Gamma(t+1)\Gamma(m-n+t+1)\Gamma(l-m-t+1)}
\sin^{2t}\frac{\theta}{2}.\label{Psin}
\end{multline}
Here $\varphi$, $\theta$, $\psi$ are real Euler parameters for
$\SU(2)$ (the first column from the relations (\ref{CEA})). At
$m\geq n$ the function $P^l_{mn}(\cos\theta)$ is expressed via the
hypergeometric function as
\begin{multline}\label{Spher1}
P^l_{mn}(\cos\theta)=\frac{i^{m-n}}{\Gamma(m-n+1)}\sqrt{\frac{\Gamma(l-n+1)\Gamma(l+m+1)}
{\Gamma(l-m+1)\Gamma(l+n+1)}}\times\\
\times\cos^{m+n}\frac{\theta}{2}\sin^{m-n}\frac{\theta}{2}
\hypergeom{2}{1}{l+m+1,m-l}{m-n+1}{\sin^2\frac{\theta}{2}}.
\end{multline}
Analogously, at $n\geq m$
\begin{multline}\label{Spher2}
P^l_{mn}(\cos\theta)=\frac{i^{n-m}}{\Gamma(n-m+1)}\sqrt{\frac{\Gamma(l-m+1)\Gamma(l+n+1)}
{\Gamma(l-n+1)\Gamma(l+m+1)}}\times\\
\times\cos^{m+n}\frac{\theta}{2}\sin^{n-m}\frac{\theta}{2}
\hypergeom{2}{1}{l+n+1,n-l}{n-m+1}{\sin^2\frac{\theta}{2}}.
\end{multline}
It is easy to see that the functions (\ref{Spher1}) and
(\ref{Spher2}) with an accuracy of the constant coincide with the
function (\ref{Sol1}) (correspondingly (\ref{Hyper2})) if to open
the modules and suppose $z=\cos\theta$.

Other expression for the function $P^l_{mn}(\cos\theta)$, related
with (\ref{Psin}), is defined by the transformation $u=kz$, where
$k=\ar\begin{pmatrix}
\bar{\alpha}^{-1} & \beta\\
0 & \bar{\alpha}
\end{pmatrix}$ and
$z=\ar\begin{pmatrix}
1 & 0\\
-\bar{\beta}/\bar{\alpha} & 1
\end{pmatrix}$. This expression has the form
\begin{multline}\label{Ptan}
P^l_{mn}(\cos\theta)=
i^{m-n}\sqrt{\Gamma(l-m+1)\Gamma(l+m+1)\Gamma(l-n+1)
\Gamma(l+n+1)}\times\\
\cos^{2l}\frac{\theta}{2}\tg^{m-n}\frac{\theta}{2}\times\\
\sum^{\min(l-m,l+n)}_{j=\max(0,n-m)}
\frac{i^{2j}\tg^{2j}\dfrac{\theta}{2}}
{\Gamma(j+1)\Gamma(l-m-j+1)\Gamma(l+n-j+1)\Gamma(m-n+j+1)}.
\end{multline}
Correspondingly,  the functions (\ref{Ptan}) are expressed via the
hypergeometric function as follows:
\begin{multline}
P^l_{mn}(\cos\theta)=i^{m-n}\sqrt{\frac{\Gamma(l+m+1)\Gamma(l-n+1)}
{\Gamma(l-m+1)\Gamma(l+n+1)}}\times\\
\times\cos^{2l}\frac{\theta}{2}\tg^{m-n}\frac{\theta}{2}
\hypergeom{2}{1}{m-l,-n-l}{m-n+1}{-\tg^2\frac{\theta}{2}},\quad
m\geq n; \label{Spher3}
\end{multline}
\begin{multline}
P^l_{mn}(\cos\theta)=i^{n-m}\sqrt{\frac{\Gamma(l+n+1)\Gamma(l-m+1)}
{\Gamma(l-n+1)\Gamma(l+m+1)}}\times\\
\times\cos^{2l}\frac{\theta}{2}\tg^{n-m}\frac{\theta}{2}
\hypergeom{2}{1}{n-l,-m-l}{n-m+1}{-\tg^2\frac{\theta}{2}},\quad
n\geq m. \label{Spher4}
\end{multline}

In turn, matrix elements of irreducible representations of the group
$\SU(1,1)$ have the form \cite{Vil65,VK90}
\[
t^l_{mn}(g)=e^{-im\epsilon}\fP^l_{mn}(\ch\tau)e^{-in\varepsilon},
\]
where in the case of finite-dimensional representations
\begin{multline}
\fP^l_{mn}(\ch\tau)=
\sqrt{\frac{\Gamma(l-m+1)\Gamma(l-n+1)}{\Gamma(l+m+1)\Gamma(l+n+1)}}\times\\
\times\ch^{m+n}\frac{\theta}{2}\sh^{m+n}\frac{\theta}{2}\sum^{l-m}_{s=0}
\frac{(-1)^s\Gamma(l+m+s+1)}{\Gamma(s+1)\Gamma(m-n+s+1)\Gamma(l-m-s+1)}
\sh^{2s}\frac{\theta}{2},\label{Bsin}
\end{multline}
or
\begin{multline}\label{Btan}
\fP^l_{mn}(\ch\tau)= \sqrt{\Gamma(l-m+1)\Gamma(l+m+1)\Gamma(l-n+1)
\Gamma(l+n+1)}\times\\
\ch^{2l}\frac{\tau}{2}\tnh^{m-n}\frac{\tau}{2}\times\\
\sum^{\min(l-m,l+n)}_{s=\max(0,n-m)} \frac{\tnh^{2s}\dfrac{\tau}{2}}
{\Gamma(s+1)\Gamma(l-m-s+1)\Gamma(l+n-s+1)\Gamma(m-n+s+1)}.
\end{multline}
Here $\epsilon$, $\tau$, $\varepsilon$ are real Euler parameters for
the group $\SU(1,1)$ (the second column from (\ref{CEA}) at the
restriction of the parameters $\epsilon$ and $\varepsilon$ within
the limits $0\leq\epsilon\leq 2\pi$ and $-2\pi\leq\varepsilon <
2\pi$). The functions $\fP^l_{mn}(\ch\tau)$ can be reduced also to
hypergeometric functions. So, at $m\geq n$ we have
\begin{multline}
\fP^l_{mn}(\ch\tau)=
\frac{1}{\Gamma(m-n+1)}\sqrt{\frac{\Gamma(l-n+1)\Gamma(l+m+1)}
{\Gamma(l-m+1)\Gamma(l+n+1)}}\times\\
\times\ch^{m+n}\frac{\tau}{2}\sh^{m-n}\frac{\tau}{2}
\hypergeom{2}{1}{l+m+1,m-l}{m-n+1}{-\sh^2\frac{\tau}{2}}=\\
=\sqrt{\frac{\Gamma(l+m+1)\Gamma(l-n+1)}
{\Gamma(l-m+1)\Gamma(l+n+1)}}
\ch^{2l}\frac{\tau}{2}\tnh^{m-n}\frac{\tau}{2}
\hypergeom{2}{1}{m-l,-n-l}{m-n+1}{\tnh^2\frac{\tau}{2}}.
\label{HSpher1}
\end{multline}
Correspondingly, at $n\geq m$
\begin{multline}
\fP^l_{mn}(\ch\tau)=
\frac{1}{\Gamma(n-m+1)}\sqrt{\frac{\Gamma(l-m+1)\Gamma(l+n+1)}
{\Gamma(l-n+1)\Gamma(l+m+1)}}\times\\
\times\ch^{m+n}\frac{\tau}{2}\sh^{n-m}\frac{\tau}{2}
\hypergeom{2}{1}{l+n+1,n-l}{n-m+1}{-\sh^2\frac{\tau}{2}}=\\
=\sqrt{\frac{\Gamma(l+n+1)\Gamma(l-m+1)}
{\Gamma(l-m+1)\Gamma(l+m+1)}}
\ch^{2l}\frac{\tau}{2}\tnh^{n-m}\frac{\tau}{2}
\hypergeom{2}{1}{n-l,-m-l}{n-m+1}{\tnh^2\frac{\tau}{2}}.
\label{HSpher2}
\end{multline}
In the case of principal series of unitary representations, matrix
elements are (see, for example, \cite{Vil65,HB66})
\begin{multline}
\fP^{-\frac{1}{2}+i\rho}_{mn}(\ch\tau)=
\sqrt{\Gamma(i\rho-n+\tfrac{1}{2})\Gamma(i\rho+n+\tfrac{1}{2})
\Gamma(i\rho-m+\tfrac{1}{2})
\Gamma(i\rho+m+\tfrac{1}{2})}\times\\
\ch^{2i\rho-1}\frac{\tau}{2}\tnh^{n-m}\frac{\tau}{2}\times\\
\sum^{\infty}_{s=\max(0,m-n)}
\frac{\tnh^{2s}\frac{\tau}{2}}{\Gamma(s+1)
\Gamma(i\rho-n-s+\tfrac{1}{2})\Gamma(n-m+s+1)
\Gamma(i\rho+m-s+\tfrac{1}{2})},\label{PBtanP}
\end{multline}
or
\begin{multline}
\fP^{-\frac{1}{2}+i\rho}_{mn}(\ch\tau)=
\sqrt{\frac{\Gamma(i\rho+m+\tfrac{1}{2})
\Gamma(i\rho-n+\tfrac{1}{2})}{\Gamma(i\rho-m+\tfrac{1}{2})
\Gamma(i\rho+n+\tfrac{1}{2})}}\times\\
\ch^{2i\rho-1}\frac{\tau}{2}\tnh^{m-n}\frac{\tau}{2}
\hypergeom{2}{1}{m-i\rho+\tfrac{1}{2},-n-i\rho+\tfrac{1}{2}}
{m-n+1}{\tnh^2\frac{\tau}{2}}. \label{PBFtanh1}
\end{multline}
at $m\geq n$ and
\begin{multline}
\fP^{-\frac{1}{2}+i\rho}_{mn}(\ch\tau)=
\sqrt{\frac{\Gamma(i\rho+n+\tfrac{1}{2})
\Gamma(i\rho-m+\tfrac{1}{2})}{\Gamma(i\rho-n+\tfrac{1}{2})
\Gamma(i\rho+m+\tfrac{1}{2})}}\times\\
\ch^{2i\rho-1}\frac{\tau}{2}\tnh^{n-m}\frac{\tau}{2}
\hypergeom{2}{1}{n-i\rho+\tfrac{1}{2},-m-i\rho+\tfrac{1}{2}}
{n-m+1}{\tnh^2\frac{\tau}{2}}. \label{PBFtanh2}
\end{multline}
at $n\geq m$.

As is known \cite{Vil65}, generalized spherical functions
$P^l_{mn}(\cos\theta)$ satisfy the following addition theorem:
\begin{equation}\label{Add1}
e^{-i(m\varphi+n\psi)}P^l_{mn}(\cos\theta)=\sum_{k=-l}^le^{-ik\varphi_2}
P^l_{mk}(\cos\theta_1)P^l_{kn}(\cos\theta_2),
\end{equation}
where the angles $\varphi$, $\psi$, $\theta$, $\theta_1$,
$\varphi_2$, $\theta_2$ are related by the formulae
\begin{eqnarray}
\cos\theta&=&\cos\theta_1\cos\theta_2-\sin\theta_1\sin\theta_2\cos\varphi_2,\label{Add2}\\
e^{i\varphi}&=&\frac{\sin\theta_1\cos\theta_2+\cos\theta_1\sin\theta_2\cos\varphi_2+
i\sin\theta_2\sin\varphi_2}{\sin\theta},\label{Add3}\\
e^{\frac{i(\varphi+\psi)}{2}}&=&\frac{\cos\frac{\theta_1}{2}\cos\frac{\theta_2}{2}
e^{i\frac{\varphi_2}{2}}-\sin\frac{\theta_1}{2}\sin\frac{\theta_2}{2}
e^{-i\frac{\varphi_2}{2}}}{\cos\frac{\theta}{2}}.\label{Add4}
\end{eqnarray}
Let $\cos(\theta-i\tau)$ and $\varphi_2=0$, then the formulae
(\ref{Add2})--(\ref{Add4}) take the form
\begin{eqnarray}
\cos\theta^c&=&\cos\theta\ch\tau+i\sin\theta\sh\tau,\nonumber\\
e^{i\varphi}&=&\frac{\sin\theta\ch\tau-i\cos\theta\sh\tau}{\sin\theta^c}=1,\nonumber\\
e^{\frac{i(\varphi+\psi)}{2}}&=&\frac{\cos\frac{\theta}{2}\ch\frac{\tau}{2}+
i\sin\frac{\theta}{2}\sh\frac{\tau}{2}}{\cos\frac{\theta^c}{2}}=1.
\nonumber
\end{eqnarray}
Hence it follows that $\varphi=\psi=0$ and formula (\ref{Add1}) can
be written as
\[
Z^l_{mn}(\cos\theta^c)=\sum^l_{k=-l}P^l_{mk}(\cos\theta)\fP^l_{kn}(\ch\tau).
\]
Therefore, using the addition theorem, we derived a new
representation for the hyperspherical function. Further, taking into
account (\ref{Ptan}) and (\ref{Btan}), we obtain an explicit
expression for $Z^l_{mn}(\cos\theta^c)$,
\begin{multline}
Z^l_{mn}(\cos\theta^c)= \sum^l_{k=-l}i^{m-k}
\sqrt{\Gamma(l-m+1)\Gamma(l+m+1)\Gamma(l-k+1)\Gamma(l+k+1)}\times\\
\cos^{2l}\frac{\theta}{2}\tg^{m-k}\frac{\theta}{2}\times\\[0.2cm]
\sum^{\min(l-m,l+k)}_{j=\max(0,k-m)}
\frac{i^{2j}\tg^{2j}\dfrac{\theta}{2}}
{\Gamma(j+1)\Gamma(l-m-j+1)\Gamma(l+k-j+1)\Gamma(m-k+j+1)}\times\\[0.2cm]
\sqrt{\Gamma(l-n+1)\Gamma(l+n+1)\Gamma(l-k+1)\Gamma(l+k+1)}
\ch^{2l}\frac{\tau}{2}\tnh^{n-k}\frac{\tau}{2}\times\\[0.2cm]
\sum^{\min(l-n,l+k)}_{s=\max(0,k-n)} \frac{\tnh^{2s}\dfrac{\tau}{2}}
{\Gamma(s+1)\Gamma(l-n-s+1)\Gamma(l+k-s+1)\Gamma(n-k+s+1)}.\label{PBtan}
\end{multline}
By way of example let us calculate matrix elements
$\fM^l_{mn}(\fg)=e^{-im\varphi^c}Z^l_{mn}(\cos\theta^c)e^{-in\psi^c}$
at $l=0,\,1/2,\,1$, where $Z^l_{mn}(\cos\theta^c)$ is defined via
(\ref{PBtan}). The matrices of finite-dimensional representations at
$l=0,\,\frac{1}{2},\,1$ have the following form:
\begin{gather}
T_0(\varphi^c,\theta^c,\psi^c)=1,\label{T0}\\[0.3cm]
T_{\frac{1}{2}}(\varphi^c,\theta^c,\psi^c)=\ar\begin{pmatrix}
\fM^{\frac{1}{2}}_{-\frac{1}{2}-\frac{1}{2}} &
\fM^{\frac{1}{2}}_{\frac{1}{2}-\frac{1}{2}}\\
\fM^{\frac{1}{2}}_{-\frac{1}{2}\frac{1}{2}} &
\fM^{\frac{1}{2}}_{\frac{1}{2}\frac{1}{2}}
\end{pmatrix}=\ar\begin{pmatrix}
e^{\frac{i}{2}\varphi^c}Z^{\frac{1}{2}}_{-\frac{1}{2}-\frac{1}{2}}e^{\frac{i}{2}\psi^c}
&
e^{\frac{i}{2}\varphi^c}Z^{\frac{1}{2}}_{-\frac{1}{2}\frac{1}{2}}e^{-\frac{i}{2}\psi^c}\\
e^{-\frac{i}{2}\varphi^c}Z^{\frac{1}{2}}_{\frac{1}{2}-\frac{1}{2}}e^{\frac{i}{2}\psi^c}
&
e^{-\frac{i}{2}\varphi^c}Z^{\frac{1}{2}}_{\frac{1}{2}\frac{1}{2}}e^{-\frac{i}{2}\psi^c}
\end{pmatrix}=\nonumber\\[0.3cm]
=\ar\begin{pmatrix}
e^{\frac{i}{2}\varphi^c}\cos\frac{\theta^c}{2}e^{\frac{i}{2}\psi^c}
&
ie^{\frac{i}{2}\varphi^c}\sin\frac{\theta^c}{2}e^{-\frac{i}{2}\psi^c}\\
ie^{-\frac{i}{2}\varphi^c}\sin\frac{\theta^c}{2}e^{\frac{i}{2}\psi^c}
&
e^{-\frac{i}{2}\varphi^c}\cos\frac{\theta^c}{2}e^{-\frac{i}{2}\psi^c}
\end{pmatrix}=\nonumber\\[0.3cm]
{\renewcommand{\arraystretch}{1.3} =\begin{pmatrix}
\left[\cos\frac{\theta}{2}\ch\frac{\tau}{2}+
i\sin\frac{\theta}{2}\sh\frac{\tau}{2}\right]
e^{\frac{\epsilon+\varepsilon+i(\varphi+\psi)}{2}} &
\left[\cos\frac{\theta}{2}\sh\frac{\tau}{2}+
i\sin\frac{\theta}{2}\ch\frac{\tau}{2}\right]
e^{\frac{\epsilon-\varepsilon+i(\varphi-\psi)}{2}} \\
\left[\cos\frac{\theta}{2}\sh\frac{\tau}{2}+
i\sin\frac{\theta}{2}\ch\frac{\tau}{2}\right]
e^{\frac{\varepsilon-\epsilon+i(\psi-\varphi)}{2}} &
\left[\cos\frac{\theta}{2}\ch\frac{\tau}{2}+
i\sin\frac{\theta}{2}\sh\frac{\tau}{2}\right]
e^{\frac{-\epsilon-\varepsilon-i(\varphi+\psi)}{2}}
\end{pmatrix}},\label{T1}
\end{gather}
\begin{gather}
T_1(\varphi^c,\theta^c,\psi^c)=\ar\begin{pmatrix}
\fM^1_{-1-1} & \fM^1_{-10} & \fM^1_{-11}\\
\fM^1_{0-1} & \fM^1_{00} & \fM^1_{01}\\
\fM^1_{1-1} & \fM^1_{10} & \fM^1_{11}
\end{pmatrix}=\ar
\begin{pmatrix}
e^{i\varphi^c}Z^1_{-1-1}e^{i\psi^c} & e^{i\varphi^c}Z^1_{-10} &
e^{i\varphi^c}
Z^1_{-11}e^{-i\psi^c}\\
Z^1_{0-1}e^{i\psi^c} & Z^1_{00} & Z^1_{01}e^{-i\psi^c}\\
e^{-i\varphi^c}Z^1_{1-1}e^{i\psi^c} & e^{-i\varphi^c}Z^1_{10} &
e^{-i\varphi^c}Z^1_{11}e^{-i\psi^c}
\end{pmatrix}=\nonumber\\[0.3cm]
=\ar\begin{pmatrix}
e^{i\varphi^c}\cos^2\frac{\theta^c}{2}e^{i\psi^c} &
\frac{i}{\sqrt{2}}e^{i\varphi^c}\sin\theta^c & -e^{i\varphi^c}
\sin^2\frac{\theta^c}{2}e^{-i\psi^c}\\
\frac{i}{\sqrt{2}}\sin\theta^ce^{i\psi^c} & \cos\theta^c &
\frac{i}{\sqrt{2}}\sin\theta^ce^{-i\psi^c}\\
-e^{-i\varphi^c}\sin^2\frac{\theta^c}{2}e^{i\psi^c} &
\frac{i}{\sqrt{2}}e^{-i\varphi^c}\sin\theta^c &
e^{-i\varphi^c}\cos^2\frac{\theta^c}{2}e^{-i\psi^c}
\end{pmatrix}=\nonumber
\end{gather}
\begin{multline}
{\renewcommand{\arraystretch}{1.1}=\left(\begin{array}{cc}\scr
\left[\cos^2\frac{\theta}{2}\ch^2\frac{\tau}{2}+\frac{i\sin\theta\sh\tau}{2}-
\sin^2\frac{\theta}{2}\sh^2\frac{\tau}{2}\right]e^{\epsilon+\varepsilon+i(\varphi+\psi)}
&\scr
\left[\frac{1}{\sqrt{2}}(\cos\theta\sh\tau+i\sin\theta\ch\tau)\right]e^{\epsilon+i\varphi} \\
\scr\left[\frac{1}{\sqrt{2}}(\cos\theta\sh\tau+i\sin\theta\ch\tau)\right]e^{\varepsilon+i\psi}
&\scr
\cos\theta\ch\tau+i\sin\theta\sh\tau \\
\scr\left[\cos^2\frac{\theta}{2}\sh^2\frac{\tau}{2}+\frac{i\sin\theta\sh\tau}{2}-
\sin^2\frac{\theta}{2}\ch^2\frac{\tau}{2}\right]e^{\varepsilon-\epsilon+i(\psi-\varphi)}
&\scr
\left[\frac{1}{\sqrt{2}}(\cos\theta\sh\tau+i\sin\theta\ch\tau)\right]e^{-\epsilon-i\varphi}
\end{array}\right.}\\
{\renewcommand{\arraystretch}{1.1}\left.\begin{array}{c}\scr
\left[\cos^2\frac{\theta}{2}\sh^2\frac{\tau}{2}+\frac{i\sin\theta\sh\tau}{2}-
\sin^2\frac{\theta}{2}\ch^2\frac{\tau}{2}\right]e^{\epsilon-\varepsilon+i(\varphi-\psi)} \\
\scr\left[\frac{1}{\sqrt{2}}(\cos\theta\sh\tau+i\sin\theta\ch\tau)\right]e^{-\varepsilon-i\psi} \\
\scr\left[\cos^2\frac{\theta}{2}\ch^2\frac{\tau}{2}+\frac{i\sin\theta\sh\tau}{2}-
\sin^2\frac{\theta}{2}\sh^2\frac{\tau}{2}\right]e^{-\epsilon-\varepsilon-i(\varphi+\psi)}
\end{array}\right).}\label{T2}
\end{multline}

Let us express now $Z^l_{mn}(\cos\theta^c)$ via the hypergeometric
functions. So, at $m\geq n$ from (\ref{Spher3}) and (\ref{HSpher1})
it follows that
\begin{multline}
Z^l_{mn}(\cos\theta^c)=\sqrt{\frac{\Gamma(l+m+1)\Gamma(l-n+1)}{\Gamma(l-m+1)\Gamma(l+n+1)}}
\cos^{2l}\frac{\theta}{2}\ch^{2l}\frac{\tau}{2}\times\\
\sum^l_{k=-l}i^{m-k}\tg^{m-k}\frac{\theta}{2}\tnh^{k-n}\frac{\tau}{2}\times\\
\times\hypergeom{2}{1}{m-l,-k-l}{m-k+1}{-\tg^2\frac{\theta}{2}}
\hypergeom{2}{1}{k-l,-n-l}{k-n+1}{\tnh^2\frac{\tau}{2}}.
\label{PBFtan1}
\end{multline}
Analogously, at $n\geq m$ from the formulae (\ref{Spher4}) and
(\ref{HSpher2}) we have
\begin{multline}
Z^l_{mn}(\cos\theta^c)=\sqrt{\frac{\Gamma(l-m+1)\Gamma(l+n+1)}{\Gamma(l+m+1)\Gamma(l-n+1)}}
\cos^{2l}\frac{\theta}{2}\ch^{2l}\frac{\tau}{2}\times\\
\sum^l_{k=-l}i^{k-m}\tg^{k-m}\frac{\theta}{2}\tnh^{n-k}\frac{\tau}{2}\times\\
\times\hypergeom{2}{1}{k-l,-m-l}{k-m+1}{-\tg^2\frac{\theta}{2}}
\hypergeom{2}{1}{n-l,-k-l}{n-k+1}{\tnh^2\frac{\tau}{2}}.
\label{PBFtan2}
\end{multline}
Other representation for $Z^l_{mn}(\cos\theta^c)$ in the form of
hypergeometric function we obtain from the formulae (\ref{Spher1}),
(\ref{Spher2}) and (\ref{HSpher1}), (\ref{HSpher2}). Namely,
\begin{multline}
Z^l_{mn}(\cos\theta^c)=\sqrt{\frac{\Gamma(l+m+1)\Gamma(l-n+1)}{\Gamma(l-m+1)\Gamma(l+n+1)}}
\sum^l_{k=-l}\frac{i^{m-k}}{\Gamma(m-k+1)\Gamma(k-n+1)}\times\\
\times\cos^{m+k}\frac{\theta}{2}\sin^{m-k}\frac{\theta}{2}\sh^{k-n}\frac{\tau}{2}
\ch^{k+n}\frac{\tau}{2}\times\\
\times\hypergeom{2}{1}{l+m+1,m-l}{m-k+1}{\sin^2\frac{\theta}{2}}
\hypergeom{2}{1}{k+l+1,k-l}{k-n+1}{-\sh^2\frac{\tau}{2}},\quad m\geq
n;\nonumber
\end{multline}
\begin{multline}
Z^l_{mn}(\cos\theta^c)=\sqrt{\frac{\Gamma(l-m+1)\Gamma(l+n+1)}{\Gamma(l+m+1)\Gamma(l-n+1)}}
\sum^l_{k=-l}\frac{i^{k-m}}{\Gamma(k-m+1)\Gamma(n-k+1)}\times\\
\times\cos^{m+k}\frac{\theta}{2}\sin^{k-m}\frac{\theta}{2}\sh^{n-k}\frac{\tau}{2}
\ch^{k+n}\frac{\tau}{2}\times\\
\times\hypergeom{2}{1}{l+k+1,k-l}{k-m+1}{\sin^2\frac{\theta}{2}}
\hypergeom{2}{1}{n+l+1,n-l}{n-k+1}{-\sh^2\frac{\tau}{2}},\quad n\geq
m.\nonumber
\end{multline}
Analogous expressions take place for the functions
$Z^{\dot{l}}_{\dot{m}\dot{n}}(\cos\dot{\theta}^c)$.

Relativistic spherical functions of the second type
$f(\varphi^c,\theta^c)=\fM^m_l(\varphi^c,\theta^c,0)=e^{-im\varphi^c}
Z^m_l(\cos\theta^c)$, where
\[
Z^m_l(\cos\theta^c)=\sum^l_{k=-l}P^l_{mk}(\cos\theta)\fP^k_l(\cosh\tau),
\]
are defined on the surface of complex two-sphere (\ref{CS}). In its
turn, the functions
$f(\dot{\varphi}^c,\dot{\theta}^c)=e^{i\dot{m}\dot{\varphi}^c}
Z^{\dot{m}}_{\dot{l}}(\cos\dot{\theta}^c)$ are defined on the
surface of dual sphere (\ref{DS}). Explicit expressions and
hypergeometric type formulae for $f(\varphi^c,\theta^c)$ and
$f(\dot{\varphi}^c,\dot{\theta}^c)$ follow directly from the
previous expressions for $f(\fg)$ at $n=0$.

\section{Relativistic spherical functions of unitary representations of $\SO_0(1,3)$}
Relativistic spherical functions
$\fM^l_{mn}(\varphi^c,\theta^c,\psi^c)$, considered in the previous
sections, define matrix elements of non-unitary finite-dimensional
representations of the group $\SO_0(1,3)$. As is known \cite{GMS},
finite-dimensional (spinor) representations of $\SO_0(1,3)$ in the
space of symmetric polynomials $\Sym_{(k,r)}$ have the following
form:
\begin{equation}\label{TenRep}
T_{\fg}q(\xi,\overline{\xi})=(\gamma\xi+\delta)^{l_0+l_1-1}
\overline{(\gamma\xi+\delta)}^{l_0-l_1+1}q\left(\frac{\alpha\xi+\beta}{\gamma\xi+\delta};
\frac{\overline{\alpha\xi+\beta}}{\overline{\gamma\xi+\delta}}\right),
\end{equation}
where $k=l_0+l_1-1$, $r=l_0-l_1+1$, and the pair $(l_0,l_1)$ defines
some representation of $\SO_0(1,3)$ in the Gel'fand-Naimark basis:
\begin{eqnarray}
H_{3}\xi_{k\nu} &=&m\xi_{k\nu},\nonumber\\
H_{+}\xi_{k\nu} &=&\sqrt{(k+\nu+1)(k-\nu)}\xi_{k,\nu+1},\nonumber\\
H_{-}\xi_{k\nu} &=&\sqrt{(k+\nu)(k-\nu+1)}\xi_{k,\nu-1},\nonumber\\
F_{3}\xi_{k\nu} &=&C_{l}\sqrt{k^{2}-\nu^{2}}\xi_{k-1,\nu}-A_{l}\nu\xi_{k,\nu}-\nonumber \\
&&\hspace{1.8cm}-C_{k+1}\sqrt{(k+1)^{2}-\nu^{2}}\xi_{k+1,\nu},\nonumber\\
F_{+}\xi_{k\nu} &=&C_{k}\sqrt{(k-\nu)(k-\nu-1)}\xi_{k-1,\nu+1}-\nonumber\\
&&\hspace{1.3cm}-A_{k}\sqrt{(k-\nu)(k+\nu+1)}\xi_{k,\nu+1}+\nonumber \\
&&\hspace{1.8cm}+C_{k+1}\sqrt{(k+\nu+1)(k+\nu+2)}\xi_{k+1,\nu+1},\nonumber\\
F_{-}\xi_{k\nu} &=&-C_{k}\sqrt{(k+\nu)(k+\nu-1)}\xi_{k-1,\nu-1}-\nonumber\\
&&\hspace{1.3cm}-A_{k}\sqrt{(k+\nu)(k-\nu+1)}\xi_{k,\nu-1}-\nonumber\\
&&\hspace{1.8cm}-C_{k+1}\sqrt{(k-\nu+1)(k-\nu+2)}\xi_{k+1,\nu-1},\nonumber
\end{eqnarray}
\begin{equation}\label{GNB}
A_{k}=\frac{il_{0}l_{1}}{k(k+1)},\quad
C_{k}=\frac{i}{k}\sqrt{\frac{(k^{2}-l^{2}_{0})(k^{2}-l^{2}_{1})}
{4k^{2}-1}},
\end{equation}
$$\nu=-k,-k+1,\ldots,k-1,k,$$
$$k=l_{0}\,,l_{0}+1,\ldots,$$
where $l_{0}$ is positive integer or half-integer number, $l_{1}$ is
an arbitrary complex number. These formulae define a
finite--dimensional representation of the group $\SO_0(1,3)$ when
$l^2_1=(l_0+p)^2$, $p$ is some natural number. In the case
$l^2_1\neq(l_0+p)^2$ we have an infinite-dimensional representation
of $\SO_0(1,3)$. The operators $H_{3},H_{+},H_{-},F_{3},F_{+},F_{-}$
are
\begin{eqnarray}
&&H_+=i\sA_1-\sA_2,\quad H_-=i\sA_1+\sA_2,\quad H_3=i\sA_3,\nonumber\\
&&F_+=i\sB_1-\sB_2,\quad F_-=i\sB_1+\sB_2,\quad F_3=i\sB_3.\nonumber
\end{eqnarray}
This basis was first given by Gel'fand in 1944 (see also
\cite{Har47,GY48,Nai58}). The following relations between generators
$\sY_\pm$, $\sX_\pm$, $\sY_3$, $\sX_3$ and $H_\pm$, $F_\pm$, $H_3$,
$F_3$ define a relationship between the Van der Waerden and
Gel'fand-Naimark bases:
\[
{\renewcommand{\arraystretch}{1.7}
\begin{array}{ccc}
\sY_+&=&-\dfrac{1}{2}(F_++iH_+),\\
\sY_-&=&-\dfrac{1}{2}(F_-+iH_-),\\
\sY_3&=&-\dfrac{1}{2}(F_3+iH_3),
\end{array}\quad
\begin{array}{ccc}
\sX_+&=&\dfrac{1}{2}(F_+-iH_+),\\
\sX_-&=&\dfrac{1}{2}(F_--iH_-),\\
\sX_3&=&\dfrac{1}{2}(F_3-iH_3).
\end{array}.
}
\]
The relation between the numbers $l_0$, $l_1$ and the number $l$
(the weight of representation in the basis (\ref{Waerden})) is given
by a following formula:
\[
(l_0,l_1)=\left(l,l+1\right),
\]
whence it immediately follows that
\begin{equation}\label{RelL}
l=\frac{l_0+l_1-1}{2}.
\end{equation}
As is known \cite{GMS}, if an irreducible representation of the
proper Lorentz group $\SO_0(1,3)$ is defined by the pair
$(l_0,l_1)$, then a conjugated representation is also irreducible
and is defined by a pair $\pm(l_0,-l_1)$. Therefore,
\[
(l_0,l_1)=\left(-\dot{l},\,\dot{l}+1\right).
\]
Hence it follows that
\begin{equation}\label{RelDL}
\dot{l}=\frac{l_0-l_1+1}{2}.
\end{equation}

For the unitary representations, that is, in the case of principal
series representations of the group $\SO_0(1,3)$, there exists an
analogue of the formula (\ref{TenRep}),
\begin{equation}\label{Principal}
V_af(z)=(a_{12}z+a_{22})^{\frac{\lambda}{2}+i\frac{\rho}{2}-1}
\overline{(a_{12}z+a_{22})}^{-\frac{\lambda}{2}+i\frac{\rho}{2}-1}
f\left(\frac{a_{11}z+a_{21}}{a_{12}z+a_{22}}\right),
\end{equation}
where $f(z)$ is a measurable function of the Hilbert space $L_2(Z)$,
satisfying the condition $\int|f(z)|^2dz<\infty$, $z=x+iy$. A
totality of all representations $a\rightarrow T^\alpha$,
corresponding to all possible pairs $\lambda$, $\rho$, is called a
principal series of representations of the group $\SO_0(1,3)$ and
denoted as $\fS_{\lambda,\rho}$. At this point, a comparison of
(\ref{Principal}) with the formula (\ref{TenRep}) for the spinor
representation $\fS_l$ shows that the both formulas have the same
structure; only the exponents at the factors $(a_{12}z+a_{22})$,
$\overline{(a_{12}z+a_{22})}$ and the functions $f(z)$ are
different. In the case of spinor representations the functions
$f(z)$ are polynomials $p(z,\bar{z})$ in the spaces $\Sym_{(k,r)}$,
and in the case of a representation $\fS_{\lambda,\rho}$ of the
principal series $f(z)$ are functions from the Hilbert space
$L_2(Z)$.

We know that a representation $S_l$ of the group $\SU(2)$ is
realized in terms of the functions
$t^l_{mn}(u)=e^{-im\varphi}P^l_{mn}(\cos\theta)e^{-in\psi}$. We use
below the following
\begin{theorem}[{\rm Naimark \cite{Nai58}}]
The representation $S_k$ of $\SU(2)$ is contained in
$\fS_{\lambda,\rho}$ not more then one time. At this point, $S_k$ is
contained in $\fS_{\lambda,\rho}$, when $\frac{\lambda}{2}$ is one
from the numbers $-k,-k+1,\ldots,k$.
\end{theorem}

Let us find a relation between the parameters $l_0$, $l_1$ in the
formulae (\ref{GNB}) and the parameters $\lambda$, $\rho$ of the
representation $\fS_{\lambda,\rho}$. The number $l_0$ is the lowest
from the weights $k$ of representations $S_k$ contained in
$\fS_{\lambda,\rho}$. Hence it follows that
$k\geq\left|\frac{\lambda}{2}\right|$ and
$l_0=\left|\frac{\lambda}{2}\right|$. Let us consider the operators
\begin{equation}\label{Cas1}
\Delta=F_+F_-+F_-F_++2F^2_3-(H_+H_-+H_-H_++2H^2_3)=\sA^2-\sB^2,
\end{equation}
\begin{equation}\label{Cas2}
\Delta^\prime=H_+F_-+H_-F_++F_+H_-+F_-H_++4H_3F_3=\sA\sB.
\end{equation}
Applying the formulae (\ref{GNB}), we obtain
\begin{eqnarray}
\Delta\xi_{k\nu}&=&-2(l^2_0+l^2_1-1)\xi_{k\nu},\label{Cas3}\\
\Delta^\prime\xi_{k\nu}&=&-4il_0l_1\xi_{k\nu}.\label{Cas4}
\end{eqnarray}
On the other hand, calculating infinitesimal operators of the
representation $\fS_{\lambda,\rho}$ in the space $L_2(Z)$, we have
\begin{equation}\label{Inf1}
\sA_1f=\frac{i}{2}(1-z^2)\frac{\partial f}{\partial z}-
\frac{i}{2}(1-\bar{z}^2)\frac{\partial f}{\partial\bar{z}}+
\frac{i}{2}\left[\left(\frac{\lambda}{2}+i\frac{\rho}{2}-1\right)z-
\left(-\frac{\lambda}{2}+i\frac{\rho}{2}-1\right)\bar{z}\right]f,
\end{equation}
\begin{equation}\label{Inf2}
\sA_2f=\frac{1}{2}(1+z^2)\frac{\partial f}{\partial z}+
\frac{1}{2}(1+\bar{z}^2)\frac{\partial f}{\partial\bar{z}}-
\frac{1}{2}\left[\left(\frac{\lambda}{2}+i\frac{\rho}{2}-1\right)z+
\left(-\frac{\lambda}{2}+i\frac{\rho}{2}-1\right)\bar{z}\right]f,
\end{equation}
\begin{equation}\label{Inf3}
\sA_3f=iz\frac{\partial f}{\partial z}-i\bar{z}\frac{\partial
f}{\partial\bar{z}}-\frac{i}{2}\lambda f,
\end{equation}
\begin{equation}\label{Inf4}
\sB_1f=\frac{1}{2}(1-z^2)\frac{\partial f}{\partial z}+
\frac{1}{2}(1-\bar{z}^2)\frac{\partial f}{\partial\bar{z}}+
\frac{1}{2}\left[\left(\frac{\lambda}{2}+i\frac{\rho}{2}-1\right)z+
\left(-\frac{\lambda}{2}+i\frac{\rho}{2}-1\right)\bar{z}\right]f,
\end{equation}
\begin{equation}\label{Inf5}
\sB_2f=-\frac{i}{2}(1+z^2)\frac{\partial f}{\partial z}+
\frac{i}{2}(1+\bar{z}^2)\frac{\partial f}{\partial\bar{z}}+
\frac{i}{2}\left[\left(\frac{\lambda}{2}+i\frac{\rho}{2}-1\right)z-
\left(-\frac{\lambda}{2}+i\frac{\rho}{2}-1\right)\bar{z}\right]f,
\end{equation}
\begin{equation}\label{Inf6}
\sB_3f=z\frac{\partial f}{\partial z}+\bar{z}\frac{\partial
f}{\partial\bar{z}}+\left(1-i\frac{\rho}{2}\right)f,
\end{equation}
Substituting the latter expressions into (\ref{Cas1}) and
(\ref{Cas2}), we find that
\begin{eqnarray}
\Delta
f(z)&=&-2\left[\left(\frac{\lambda}{2}\right)^2-\left(\frac{\rho}{2}\right)^2-1
\right]f(z),\nonumber\\
\Delta^\prime f(z)&=&-\lambda\rho f(z).\nonumber
\end{eqnarray}
The comparison of these formulae with (\ref{Cas3}) and (\ref{Cas4})
shows that
\begin{eqnarray}
&&\left(\frac{\lambda}{2}\right)^2-\left(\frac{\rho}{2}\right)^2=l^2_0+l^2_1,\label{Cas5}\\
&&\lambda\rho=4i l_0l_1.\label{Cas6}
\end{eqnarray}
Let $l_0\neq 0$; since $l_0=\left|\frac{\lambda}{2}\right|$, then
from (\ref{Cas6}) it follows that
\[
l_1=-i(\sign\lambda)\frac{\rho}{2}.
\]
If $l_0=0$ and, therefore, $\lambda=0$, from (\ref{Cas5}) we obtain
\[
l_1=\pm i\frac{\rho}{2}.
\]
Thus, the numbers $l_0$, $l_1$ è $\lambda$, $\rho$ are related by
the formulas
\begin{gather}
l_0=\left|\frac{\lambda}{2}\right|,\quad
l_1=-i(\sign\lambda)\frac{\rho}{2}\quad\text{if $\lambda\neq 0$},\nonumber\\
l_0=0,\quad l_1=\pm i\frac{\rho}{2}\quad\text{if
$\lambda=0$}.\nonumber
\end{gather}

On the other hand, we see from (\ref{KO}) that Laplace-Beltrami
operators $\sX^2=-l(l+1)$ and $\sY^2=-\dot{l}(\dot{l}+1)$ contain
Casimir operators $\Delta=\sA^2-\sB^2$ and
$\Delta^\prime=\sA\cdot\sB$ as real and imaginary parts:
\begin{eqnarray}
\sX^2&=&\frac{1}{4}\Delta+\frac{1}{2}i\Delta^\prime=-l(l+1),\nonumber\\
\sY^2&=&\frac{1}{4}\Delta-\frac{1}{2}i\Delta^\prime=-\dot{l}(\dot{l}+1).\nonumber
\end{eqnarray}
Taking into account in the latter operators the formulae
(\ref{Cas3}) and (\ref{Cas4}), we arrive at relations (\ref{RelL})
and (\ref{RelDL}). The relations (\ref{RelL}) and (\ref{RelDL})
define a relation between parameters $l_0$, $l_1$ of the
Gel'fand-Naimark basis (\ref{GNB}) and parameters $l$, $\dot{l}$ of
the Van der Waerden basis (\ref{Waerden}).

As is known, all the unitary representations of the group
$\SO_0(1,3)$ are infinite-dimensional. The group $\SO_0(1,3)$ is
non-compact and one from its real forms, the group $\SU(1,1)$, is
also non-compact group involving unitary infinite-dimensional
representations. In the previous section it has been shown that the
matrix elements of $\SO_0(1,3)$  are defined via the addition
theorem for the matrix elements of the subgroups $\SU(2)$ and
$\SU(1,1)$. This factorization allows us to separate explicitly  in
the matrix element all the parameters changing in infinite limits.

In such a way, using Theorem 1, formulae (\ref{PBtanP}),
(\ref{PBtan}) and (\ref{RelL}), we find that matrix elements of the
principal series representations of the group $\SO_0(1,3)$ have the
form
\begin{multline}
\fM^{-\frac{1}{2}+i\rho,l_0}_{mn}(\mathfrak{g})=
e^{-m(\epsilon+i\varphi)-n(\varepsilon+i\psi)}
Z^{-\frac{1}{2}+i\rho,l_0}_{mn}=
e^{-m(\epsilon+i\varphi)-n(\varepsilon+i\psi)}\times\\[0.2cm]
\sum^{l_0}_{t=-l_0}i^{m-t} \sqrt{\Gamma(l_0-m+1)
\Gamma(l_0+m+1)\Gamma(l_0-t+1)
\Gamma(l_0+t+1)}\times\\
\cos^{2l_0}\frac{\theta}{2}\tg^{m-t}\frac{\theta}{2}\times\\[0.2cm]
\sum^{\min(l_0-m,l_0+t)}_{j=\max(0,t-m)}
\frac{i^{2j}\tg^{2j}\dfrac{\theta}{2}} {\Gamma(j+1)\Gamma(l_0-m-j+1)
\Gamma(l_0+t-j+1)\Gamma(m-t+j+1)}\times\\[0.2cm]
\sqrt{\Gamma(\tfrac{1}{2}+i\rho-n)
\Gamma(\tfrac{1}{2}+i\rho+n)\Gamma(\tfrac{1}{2}+i\rho-t)
\Gamma(\tfrac{1}{2}+i\rho+t)}
\ch^{-1+2i\rho}\frac{\tau}{2}\tnh^{n-t}\frac{\tau}{2}\times\\[0.2cm]
\sum^{\infty}_{s=\max(0,t-n)} \frac{\tnh^{2s}\dfrac{\tau}{2}}
{\Gamma(s+1)\Gamma(\frac{1}{2}+i\rho-n-s)
\Gamma(\frac{1}{2}+i\rho+t-s)\Gamma(n-t+s+1)},\label{MPrincip}
\end{multline}
where $l_0=\left|\frac{\lambda}{2}\right|$ and $\frac{\lambda}{2}$
is one from the numbers $-k,-k+1,\ldots,k$. It is obvious that
$\fM^{-\frac{1}{2}+i\rho,l_0}_{mn}(\fg)$ cannot be attributed as
matrix elements to single irreducible representation. From the
latter expression it follows that relativistic spherical functions
$f(\fg)$ of the principal series can be defined by means of the
function
\begin{equation}\label{MEPS}
\fM^{-\frac{1}{2}+i\rho,l_0}_{mn}(\fg)=e^{-m(\epsilon+i\varphi)}
Z^{-\frac{1}{2}+i\rho,l_0}_{mn}(\cos\theta^c)e^{-n(\varepsilon+i\psi)},
\end{equation}
where
\[
Z^{-\frac{1}{2}+i\rho,l_0}_{mn}(\cos\theta^c)= \sum^{l_0}_{t=-l_0}
P^{l_0}_{mt}(\cos\theta)\fP^{-\frac{1}{2}+i\rho}_{tn}(\ch\tau).
\]
In the case of relativistic spherical functions
$f(\varphi^c,\theta^c)$ we have
\begin{equation}\label{AHPS}
Z^m_{-\frac{1}{2}+i\rho,l_0}(\cos\theta^c)= \sum^{l_0}_{t=-l_0}
P^{l_0}_{mt}(\cos\theta)\fP^t_{-\frac{1}{2}+i\rho}(\ch\tau),
\end{equation}
where $\fP^t_{-\frac{1}{2}+i\rho}(\ch\tau)$ are {\it conical
functions} (see \cite{Bat}). In this case our result agrees with the
paper \cite{FNR66}, where matrix elements (eigenfunctions of Casimir
operators) of non-compact rotation groups are expressed in terms of
conical and spherical functions (see also \cite{Vil65}).

When $\rho$ is a cleanly imaginary number, $\rho=i\sigma$, we have
\[
T^\alpha f(z)=|a_{12}z+a_{22}|^{-2-\sigma}f
\left(\frac{a_{11}z+a_{21}}{a_{12}z+a_{22}}\right).
\]
This formula defines an unitary representation $a\rightarrow
T^\alpha$ of supplementary series $\fD_\sigma$ of the group
$\SO_0(1,3)$. In its turn, for the supplementary series $\fD_\sigma$
the following theorem holds.
\begin{theorem}[{\rm Naimark \cite{Nai58}}]
The representation $S_k$ of $\SU(2)$ is contained in
$\fS_{\lambda,\rho}$ when $k$ is an integer number. In this case,
$S_k$ is contained in $\fD_\sigma$ exactly one time.
\end{theorem}
We see that $\frac{\lambda}{2}=0$ should be one from the numbers
$-k,-k+1,\ldots,k$, therefore, when $k$ is integer.

Let us find a relation between the parameters $l_0$, $l_1$ in
(\ref{GNB}) and the parameter $\sigma$ of $\fD_\sigma$. First, the
lowest wight $l_0$ from the weights $k$ of the representations
$S_k$, contained in $\fD_\sigma$, is equal to zero, that is,
$l_0=0$. With the aim to define the parameter $l_1$ let us consider
again the Casimir operator $\Delta=\sA^2-\sB^2$. Calculating this
operator with the help of formulae (\ref{GNB}) and
(\ref{Inf1})--(\ref{Inf6}), where
$\frac{\lambda}{2}+i\frac{\rho}{2}-1$ and
$-\frac{\lambda}{2}+i\frac{\rho}{2}-1$ should be replaced by
$-\frac{\sigma}{2}-1$ and $-\frac{\sigma}{2}-1$, we obtain
\[
\Delta\xi_{k\nu}=-2(l^2_1-1)\xi_{k\nu},\quad \Delta
f(z)=-2\left[\left(\frac{\sigma}{2}\right)^2-1\right]f(z).
\]
Hence it follows that $l^2_1=\left(\frac{\sigma}{2}\right)^2$ and
$l_1=\pm\frac{\sigma}{2}$ (the choice of the sign is not important).
Thus, for the supplementary series the relations
\[
l_0=0,\quad l_1=\pm\frac{\sigma}{2}
\]
hold. In the case of $\fD_\sigma$, Laplace-Beltrami operators
$\sX^2$ and $\sY^2$ are coincide with each other,
$\sX^2=\sY^2=\sA^2-\sB^2$. This means that we come here to
representations of $\SO_0(1,3)$ restricted to the subgroup
$\SU(1,1)$.

Thus, matrix elements of supplementary series appear as a particular
case of the matrix elements of the principal series at $l_0=0$ and
$\rho=i\sigma$:
\begin{multline}
\fM^{-\frac{1}{2}-\sigma}_{mn}(\mathfrak{g})=
e^{-m(\epsilon+i\varphi)-n(\varepsilon+i\psi)}
Z^{-\frac{1}{2}-\sigma}_{mn}=
e^{-m(\epsilon+i\varphi)-n(\varepsilon+i\psi)}\times\\[0.2cm]
\sqrt{\Gamma(\tfrac{1}{2}-\sigma-n)
\Gamma(\tfrac{1}{2}-\sigma+n)\Gamma(\tfrac{1}{2}-\sigma-m)
\Gamma(\tfrac{1}{2}-\sigma+m)}
\ch^{-1-2\sigma}\frac{\tau}{2}\tnh^{n-m}\frac{\tau}{2}\times\\[0.2cm]
\sum^{\infty}_{s=\max(0,m-n)} \frac{\tnh^{2s}\dfrac{\tau}{2}}
{\Gamma(s+1)\Gamma(\tfrac{1}{2}-\sigma-n-s)
\Gamma(\tfrac{1}{2}-\sigma+m-s)\Gamma(n-m+s+1)}.\label{MSupl}
\end{multline}
Or
\[
\fM^{-\frac{1}{2}-\sigma}_{mn}(\fg)=e^{-m(\epsilon+i\varphi)}
\fP^{-\frac{1}{2}-\sigma}_{mn}(\ch\tau)e^{-n(\varepsilon+i\psi)},
\]
that is, the hyperspherical function
$Z^{-\frac{1}{2}+i\rho,l_0}_{mn}(\cos\theta^c)$ in the case of
supplementary series is reduced to the Jacobi function
$\fP^{-\frac{1}{2}-\sigma}_{mn}(\ch\tau)$. For the relativistic
spherical functions $f(\varphi^c,\theta^c)\sim f(\varphi^c,\tau)$ of
supplementary series we obtain
\[
\fM^m_{-\frac{1}{2}-\sigma}(\fg)=e^{-m(\epsilon+i\varphi)}
\fP^m_{-\frac{1}{2}-\sigma}(\ch\tau).
\]

Let us express now the relativistic spherical function
$\fM^{-\frac{1}{2}+i\rho}_{mn}(\fg)$ of the principal series
representations of $\SO_0(1,3)$ via the hypergeometric function.
Using the formulae (\ref{MPrincip}), (\ref{PBFtan1}),
(\ref{PBFtan2}) and (\ref{PBFtanh1}), (\ref{PBFtanh2}), we find
\begin{multline}
\fM^{-\frac{1}{2}+i\rho,l_0}_{mn}(\fg)=e^{-m(\epsilon+i\varphi)-n(\varepsilon+i\psi)}
\sqrt{\frac{\Gamma(l_0+m+1)\Gamma(i\rho-n+\frac{1}{2})}
{\Gamma(l_0-m+1)\Gamma(i\rho+n+\frac{1}{2})}}\times\\
\times\cos^{2l_0}\frac{\theta}{2}\ch^{-1+2i\rho}\frac{\tau}{2}
\sum^{l_0}_{t=-l_0}
i^{m-t}\tg^{m-t}\frac{\theta}{2}\tnh^{t-n}\frac{\tau}{2}\times\\
\times\hypergeom{2}{1}{m-l_0,-t-l_0}{m-t+1}{-\tg^2\frac{\theta}{2}}
\hypergeom{2}{1}{t-i\rho+\frac{1}{2},-n-i\rho+\frac{1}{2}}{t-n+1}{\tnh^2\frac{\tau}{2}},
\quad m\geq n; \nonumber
\end{multline}
\begin{multline}
\fM^{-\frac{1}{2}+i\rho,l_0}_{mn}(\fg)=e^{-m(\epsilon+i\varphi)-n(\varepsilon+i\psi)}
\sqrt{\frac{\Gamma(l_0-m+1)\Gamma(i\rho+n+\frac{1}{2})}
{\Gamma(l_0+m+1)\Gamma(i\rho-n+\frac{1}{2})}}\times\\
\times\cos^{2l_0}\frac{\theta}{2}\ch^{-1+2i\rho}\frac{\tau}{2}
\sum^{l_0}_{t=-l_0}
i^{t-m}\tg^{t-m}\frac{\theta}{2}\tnh^{n-t}\frac{\tau}{2}\times\\
\times\hypergeom{2}{1}{t-l_0,-m-l_0}{t-m+1}{-\tg^2\frac{\theta}{2}}
\hypergeom{2}{1}{n-i\rho+\frac{1}{2},-t-i\rho+\frac{1}{2}}{n-t+1}{\tnh^2\frac{\tau}{2}},
\quad n\geq m. \nonumber
\end{multline}
For the supplementary series we have
\begin{multline}
\fM^{-\frac{1}{2}-\sigma}_{mn}(\fg)=e^{-m(\epsilon+i\varphi)-n(\varepsilon+i\psi)}
\sqrt{\frac{\Gamma(m-\sigma+\frac{1}{2})\Gamma(-n-\sigma+\frac{1}{2})}
{\Gamma(-m-\sigma+\frac{1}{2})\Gamma(n-\sigma+\frac{1}{2})}}\times\\
\times\ch^{-1-2\sigma}\frac{\tau}{2}\tnh^{m-n}\frac{\tau}{2}
\hypergeom{2}{1}{m+\sigma+\frac{1}{2},-n+\sigma+\frac{1}{2}}{m-n+1}{\tnh^2\frac{\tau}{2}},\quad
m\geq n; \nonumber
\end{multline}
\begin{multline}
\fM^{-\frac{1}{2}-\sigma}_{mn}(\fg)=e^{-m(\epsilon+i\varphi)-n(\varepsilon+i\psi)}
\sqrt{\frac{\Gamma(n-\sigma+\frac{1}{2})\Gamma(-m-\sigma+\frac{1}{2})}
{\Gamma(-n-\sigma+\frac{1}{2})\Gamma(m-\sigma+\frac{1}{2})}}\times\\
\times\ch^{-1-2\sigma}\frac{\tau}{2}\tnh^{n-m}\frac{\tau}{2}
\hypergeom{2}{1}{n+\sigma+\frac{1}{2},-m+\sigma+\frac{1}{2}}{n-m+1}{\tnh^2\frac{\tau}{2}},\quad
n\geq m. \nonumber
\end{multline}

In like manner we can define conjugated spherical functions
$f(\fg)=\fM^{-\frac{1}{2}-i\rho,l_0}_{mn}(\fg)$ and
$f(\dot{\varphi}^c,\dot{\theta}^c)=
\fM^m_{-\frac{1}{2}-i\rho,l_0}(\dot{\varphi}^c,\dot{\theta}^c,0)$ of
the principal series $\fS_{\lambda,\rho}$, since a conjugated
representation of $\SO_0(1,3)$ is defined by the pair
$\pm(l_0,-l_1)$. It is obvious that in the case of supplementary
series $\fD_\sigma$ we arrive at the same functions
$\fM^{-\frac{1}{2}-\sigma}_{mn}(\fg)$.


\begin{thebibliography}{00}
\bibitem{Dol56} A. Z. Dolginov, ``Relativistic spherical functions,"
Zh. Eksp. Teor. Fiz. {\bf 30}, 746--755 (1956).
\bibitem{DT59} A. Z. Dolginov, I. N. Toptygin, ``Relativistic spherical
functions. II," Zh. Eksp. Teor. Fiz. {\bf 37}, 1441--1451 (1959).
\bibitem{DM59} A. Z. Dolginov, A. N. Moskalev, ``Relativistic spherical
functions. III," Zh. Eksp. Teor. Fiz. {\bf 37}, 1697--1707 (1959).
\bibitem{Esk59} L. D. Eskin, ``On the theory of relativistic spherical
functions," Nauchn. dokl. vys. shkoly {\bf 2}, 95--97 (1959).
\bibitem{Foc35} V. A. Fock, ``Zur Theorie des Wasserstoffatoms,"
Zs. f. Phys. {\bf 98}, 145--154 (1935).
\bibitem{Esk61} L. D. Eskin, ``On the matrix elements of irreducible
representations of the Lorentz group," Izvestia vuzov, Matem. {\bf
6}, 179--184 (1961).
\bibitem{Gol61} V. Ya. Golodets, ``Matrix elements of irreducible
unitary and spinor representations of the homogeneous Lorentz
group," Vesti Akademii Navuk Belarusskoi SSR {\bf 1}, 19--28 (1961).
\bibitem{Str65} S. Str\"{o}m, ``On the matrix elements of a unitary
representation of the homogeneous Lorentz group," Arkiv f\"{o}r
Fysik {\bf 29}, 467--483 (1965).
\bibitem{Str67} S. Str\"{o}m, ``A note on the matrix elements of a
unitary representation of the homogeneous Lorentz group," Arkiv
f\"{o}r Fysik {\bf 33}, 465--469 (1967).
\bibitem{Str68} S. Str\"{o}m, ``Matrix elements of the supplementary
series of unitary representations of $\SL(2,C)$," Arkiv f\"{o}r
Fysik {\bf 38}, 373--381 (1968).
\bibitem{ST67} A. Sciarrino, M. Toller, ``Decomposition of the Unitary
Irreducible Representations of the Group $\SL(2,C)$ Restricted to
the Subgroup $\SU(1,1)$," J. Math. Phys. {\bf 8}, 1252--1265 (1967).
\bibitem{VD67} I. A. Verdiev, L. A. Dadashev, ``Matrix elements of the
Lorentz group unitary representation," Yad. Fiz. {\bf 6}, 1094--1099
(1967).
\bibitem{Kol70} V. I. Kolomytsev, ``The reduction of the irreducible
unitary representations of the group $\SL(2,C)$ restricted to the
subgroup $\SU(1,1)$. The additional series," Teor. Mat. Fiz. {\bf
2}, 210--229 (1970).
\bibitem{GMS} I. M. Gel'fand, R. A. Minlos, Z. Ya. Shapiro, {\em Representations
of the Rotation and Lorentz Groups and their Applications} (Pergamon
Press, Oxford, 1963).
\bibitem{Nai58} M. A. Naimark,
{\em Linear Representations of the Lorentz Group} (Pergamon, London,
1964).
\bibitem{SH70} Ya.A. Smorodinsky, M. Huszar, ``Representations of the
Lorentz group and the generalization of helicity states," Teor. Mat.
Fiz. {\bf 4}, 3, 328--340 (1970).
\bibitem{Hus85} M. Huszar, ``Spherical functions of the Lorentz group on
the hyperboloids," Acta Phys. Hung. {\bf 58}, 175--185 (1985).
\bibitem{Hus88} M. Huszar, ``Addition theorems for the spherical functions
of the Lorentz group," Acta Phys. Hung. {\bf 64}, 361--378 (1988).
\bibitem{AG64} M. Andrews, J. Gunson, ``Complex Angular momenta
and Many-Particle States. I. Local Representations of the Rotation
Group," J. Math. Phys. {\bf 5}, 1391--1400 (1964).
\bibitem{GS52} I.M. Gel'fand, Z.Ya. Shapiro, ``Representations of
the rotation group of three--dimensional space and their
applications," Uspekhi Mat. Nauk {\bf 7}, 3--117 (1952).
\bibitem{Vil65} N. Ya. Vilenkin, {\em Special Functions and the Theory of Group
Representations} (AMS, Providence, 1968).
\bibitem{Bar47} V. Bargmann, ``Irreducible unitary representations of
the Lorentz group," Ann. of Math. {\bf 48}, 568--640 (1947).
\bibitem{Kih70} A. Kihlberg, ``Fields on a homogeneous space of the
Poincar\'{e} group," Ann. Inst. Henri Poincar\'{e} {\bf 13}, 57--76
(1970).
\bibitem{Tol96} M. Toller, ``Free quantum fields on the Poincar\'{e}
group," J. Math. Phys. {\bf 37}, 2694--2730 (1996).
\bibitem{Dre97} W. Drechsler, ``Geometro-stohastically quantized fields
with internal spin variables," J. Math. Phys. {\bf 38}, 5531--5558
(1997).
\bibitem{GS01} D. M. Gitman, A. L. Shelepin, ``Fields on the Poincar\'{e} Group:
Arbitrary Spin Description and Relativistic Wave Equations," Int. J.
Theor. Phys. {\bf 40}(3), 603--684 (2001).
\bibitem{Var03c} V. V. Varlamov, ``Relativistic wavefunctions on the
Poincar\'{e} group," J. Phys. A: Math. Gen. {\bf 37}, 5467--5476
(2004).
\bibitem{Var03d} V. V. Varlamov, ``Maxwell field on the Poincar\'{e}
group," Int. J. Mod. Phys. {\bf A20}, 4095--4112 (2005).
\bibitem{AAV69} A. K. Agamaliev, N. M. Atakishiev, I. A. Verdiev,
``Invariant Expansion of Solutions of Relativistic Equations," Yad.
Fiz. {\bf 9}, 201--211 (1969).
\bibitem{Var03} V. V. Varlamov, ``General Solutions of Relativistic
Wave Equations," Int. J. Theor. Phys. {\bf 42}, No. 3, 583--633
(2003).
\bibitem{BBTD88} L. C. Biedenharn, H. W. Braden, P. Truini, H. van Dam,
``Relativistic wavefunctions on spinor spaces," J. Phys. A: Math.
Gen. {\bf 21}, 3593--3610 (1988).
\bibitem{Wa32} B. L. van der Waerden, {\em Die Gruppentheoretische Methode in der
Quantenmechanik} (Springer, Berlin, 1932).
\bibitem{VK90} N. Ya. Vilenkin, A. U. Klimyk,
{\em Representations of Lie Groups and Special Functions}, vols.
1--3. (Dordrecht: Kluwer Acad. Publ., 1991--1993).
\bibitem{HB66} W. J. Hollman, L. C. Biedenharn, ``Complex Angular
Momenta and the Groups $\SU(1,1)$ and $\SU(2)$," Annals of Physics
{\bf 39}, 1-42 (1966).
\bibitem{Har47} Harish Chandra, ``Infinite irreducible representations
of the Lorentz group," Proc. Royal. Soc. London {\bf A189}, 372--401
(1947).
\bibitem{GY48} I.M. Gel'fand, A.M. Yaglom, ``General
relativistic-invariant equations and infinite-dimensional
representations of the Lorentz group," Zh. Eksp. Teor. Fiz. {\bf
18}, 703--733 (1948).
\bibitem{Bat} H. Bateman, A. Erd\'{e}lyi, {\em Higher Transcendental Functions},
vol. I (Mc Grow-Hill Book Company, New York, 1953).
\bibitem{FNR66} J. Fischer, J. Niederle, R. Raczka, ``Generalized
Spherical Functions for the Noncompact Rotation Groups," J. Math.
Phys. {\bf 7}, 816--821 (1966).

\end{thebibliography}
\end{document}